\documentclass[a4paper]{jpconf}
\usepackage{amsmath,epsfig,graphicx}

\newcommand\as{\alpha_{\mathrm{S}}} 
\newcommand\f[2]{\frac{#1}{#2}} 
\def\beq{\begin{equation}} 
\def\eeq{\end{equation}} 
\def\beeq{\begin{eqnarray}} 
\def\eeeq{\end{eqnarray}} 
\def\to{\rightarrow}
 
\def\nn{\nonumber} 

\def\qt{q_T}
\def\mures{\mu_{res}}

\def\mgg{M_{\gamma \gamma}}
\def\Mgg{M_{\gamma \gamma}}
\def\gg{\gamma \gamma}
\def\Dpgg{\Delta \Phi_{\gamma\gamma} }

\begin{document}

\title{Transverse-momentum resummation for photon pair production at NNLL+NLO}

\author{Leandro Cieri}

\address{Physik-Institut, Universit\"at Z\"urich, CH-8057 Z\"urich, Switzerland}

\ead{lcieri@physik.uzh.ch}

\begin{abstract}
We are interested in the transverse-momentum ($q_T$) distribution of a diphoton pair produced in hadron collisions. 
We resum the logarithmically-enhanced perturbative QCD contributions at small values of $q_T$ up to next-to-next-to-leading logarithmic accuracy. We consistently combine resummation with the known next-to-leading order perturbative result at intermediate and large values of $q_T$.
We include all perturbative terms up to order $\as^2$ in our computation which, after integration over $q_T$, reproduces the known next-to-next-to-leading order result for the diphoton pair production 
total cross section. 
A comparison with LHC data is presented. We estimate the perturbative accuracy of the theoretical calculation by performing the corresponding variation of scales. We anticipate that the effect of the transverse momentum resummation is not only to recover the predictivity of the calculation at small $q_T$, but also to improve substantially the agreement with the experimental data.

\end{abstract}

\section{Introduction}
The production of two isolated photons at hadronic colliders is a very relevant process, both from the point of view of
testing the Standard Model (SM) predictions~\cite{Chatrchyan:2011qt,Aaltonen:2011vk,Chatrchyan:2014fsa,Abazov:2013pua,Aad:2011mh,Aad:2012tba} as for new physics searches.

\emph{Direct} or \emph{prompt} photons provide an ideal test to the understanding of pQCD since they constitute a theoretically and experimentally
\emph{clean} final state. From the theory side, because they do not have QCD interactions with other final state particles and, 
from the experimental point of view, because their energies and momenta can be measured with high precision in modern electromagnetic calorimeters.

Diphoton final states have played a crucial role in the recent discovery of a new boson at
the LHC \cite{cha:2012gu,aad:2012gk}, whose properties are compatible with those of the SM Higgs. They are also important in many
new physics scenarios \cite{:2012afa,:2012mx}, in particular in the search for extra-dimensions~\cite{Aad:2012cy} or supersymmetry~\cite{CMS:2012un}. 

We are interested in the process $pp\rightarrow \gamma \gamma X$, and in particular in the transverse-momentum ($q_T$) spectrum of the diphoton pair.
At the lowest-order ($\mathcal{O}(\alpha_S^0)$) the diphoton final state occurs \textit{via} the quark annihilation subprocess $q\bar{q}\rightarrow \gamma\gamma$.
The QCD corrections at the first order in the strong coupling $\alpha_S$  involve 
quark annihilation  and a new partonic channel, \textit{via} the subprocess $qg \rightarrow \gamma \gamma q$.

Several fully-differential Monte Carlo codes implement the first order corrections~\cite{Binoth:1999qq,Bern:2002jx,Campbell:2011bn,Balazs:2007hr}.
At the second order in the strong coupling $\alpha_S$ the $gg$ channel starts to contribute, and the large gluon--gluon luminosity makes
this channel potentially sizeable.

The matrix elements needed to evaluate the corrections at the second order in the strong coupling $\alpha_S$, 
for diphoton production, have been presented
in \cite{Dicus:1987fk,Barger:1989yd,Bern:1994fz,Anastasiou:2002zn}, and first
put together in a complete and consistent $\mathcal{O}(\alpha_S^2)$ calculation in the \texttt{2$\gamma$NNLO} code \cite{Catani:2011qz}.
The next-order corrections to the {\it box contribution} (which are part of the N$^3$LO QCD corrections to diphoton production)
were also computed in ref.~\cite{Bern:2002jx} and found to have a moderate quantitative effect.

The calculation of the $q_T$ spectrum requires the identification of two kinematic regions. In the large-$q_T$ region ($q_T\sim \Mgg$), where the transverse momentum is of the order of the diphoton invariant
mass $\Mgg$, calculations based on the truncation of the perturbative series at a fixed order in $\as$  are theoretically justified.
In this kinematic region, the QCD radiative corrections are known up to the next-to-leading order (NLO), including the corresponding partonic
scattering amplitudes with $X=2$~partons (at the tree level \cite{Barger:1989yd}) and the partonic scattering amplitudes with 
$X=1$~parton (up to the one-loop level \cite{Bern:1994fz}). In order to have $q_T \neq 0$ for the diphoton pair, at least one additional parton is needed. The transverse momentum spectrun of the diphoton pair has been calculated in fully-differential 
Monte Carlo codes at LO~\cite{Binoth:1999qq,Bern:2002jx,Campbell:2011bn,Balazs:2007hr} and at NLO~\cite{Catani:2011qz,DelDuca:2003uz,Gehrmann:2013aga}. Recently, first calculations for diphoton production in association with two~\cite{Bern:2014vza,Gehrmann:2013bga,Badger:2013ava} and three~\cite{Badger:2013ava} jets at NLO became available.

In the small-$q_T$ region ($q_T\ll \mgg$), where the bulk of the diphoton events is produced, the convergence of the
fixed-order expansion is spoiled by the presence of large logarithmic terms, $\as^n\ln^m (\mgg^2/q_T^2)$.
These logarithmically-enhanced terms have to be resummed to all perturbative orders
\cite{Dokshitzer:hw}--\cite{Catani:2013tia} in order to obtain reliable predictions. The resummed calculation, valid at small values of $q_T$, and the fixed-order one at large
$q_T$ have then to be consistently matched to obtain a pQCD prediction for the entire range of transverse momenta.

We use the transverse-momentum resummation formalism proposed in Refs.~\cite{Catani:2000vq,Bozzi:2005wk,Bozzi:2007pn}
(see also \cite{Catani:2010pd} for processes initiated by $gg$ annihilation). 
The formalism is valid for a generic process in which a high-mass system of non strongly-interacting particles is produced 
in hadron-hadron collisions.
The method has so far been applied to the production of the Standard Model (SM) Higgs boson 
\cite{ Bozzi:2005wk, Bozzi:2007pn, Bozzi:2003jy, deFlorian:2011xf,deFlorian:2012mx}, Higgs boson production in bottom quark annihilation~\cite{Harlander:2014hya}, Higgs boson production via gluon fusion in the MSSM~\cite{Harlander:2014uea}, single vector bosons at NLL+LO \cite{Bozzi:2008bb} and at NNLL+NLO \cite{Bozzi:2010xn} with leptonic decay~\cite{Catani:2015vma} ,
$WW$ \cite{Grazzini:2005vw,Meade:2014fca} and $ZZ$ \cite{Frederix:2008vb} pairs, vector boson pair production at NNLL+NLO \cite{Grazzini:2015wpa}, slepton pairs \cite{Bozzi:2006fw}, DY lepton pairs in polarized collisions \cite{Jiro} and recently we applied it to diphoton production at NNLL+NLO~\cite{Cieri:2015rqa}.

It is necessary to notice that besides the direct photon contribution from the hard subprocess, photons can also
be produced from the fragmentation of QCD partons. Calculate the fragmentation subprocesses requires (the poorly known)
non-perturbative information, in the form of parton fragmentation functions of the photon (the complete single- and
double-fragmentation contributions are implemented in \texttt{DIPHOX} \cite{Binoth:1999qq} for diphoton production at the first order in $\as$). Moreover, including this contribution could be non trivial at NNLO since we have particles with colour in the final state, and the $q_T$ resummation formalism, at least in its original form~\footnote{The formalism is general, as long as the measured final state is composed of non strongly-interacting particles. Transverse-momentum resummation for strongly-interacting final states, such as heavy-quark production, has been developed in Refs.~\cite{Li:2013mia,Catani:2014qha}.}, has to be generalized in order to make it possible treat colour particles in the final state. 
It is important to notice that the effect of the fragmentation
contributions is sizeably reduced by the \emph{photon isolation} criteria that are necessarily
applied in hadron collider experiments to suppress the very large rreducible background (\textit{e.g.}, photons that are faked by jets
or produced by hadron decays). Two such criteria are the so-called ``standard'' cone isolation and the ``smooth''
cone isolation proposed by Frixione \cite{Frixione:1998jh}. The standard cone isolation is easily implemented in experiments,
but it only suppresses a fraction of the fragmentation contribution. In the case of the smooth cone isolation, it (formally) eliminates the entire fragmentation contribution. 
For all of the results presented in this proceeding we rely on the smooth isolation prescription, which, for the parameters used
in the experimental analysis reproduces the standard result within a $1\%$ accuracy~\cite{Butterworth:2014efa}.

This proceeding is organized as follows. In Sect.~\ref{sec:theory} we briefly review the resummation formalism of Refs.
\cite{Catani:2000vq,Bozzi:2005wk,Bozzi:2007pn}.
In Sect.~\ref{sec:results} we present numerical results and we comment on their comparison with the LHC data \cite{Aad:2012tba}.
We also study the scale dependence of our results with the purpose of estimating the corresponding perturbative uncertainty. 
In Sect.~\ref{sec:summa} we summarize our results.

\section{The transverse-momentum resummation formalism}
\label{sec:theory}

We briefly introduce the main points of the transverse-momentum resummation formalism of Refs.~
\cite{Catani:2000vq,Bozzi:2005wk,Bozzi:2007pn}, referring to the original papers for the full details.
The formalism is general, as long as the measured final state is composed of non strongly-interacting particles (transverse-momentum resummation for strongly-interacting final states, such as heavy-quark production, has been developed in Refs.~\cite{Li:2013mia,Catani:2014qha}). Here we specialize
to the case of diphoton production only for ease of reading. The inclusive hard-scattering process considered is
\begin{equation}
h_1(p_1) + h_2(p_2) \;\to\; \gamma \gamma(\Mgg,q_T,y) + X \;\;,   
\label{first}
\end{equation}
where $h_1$ and $h_2$ are the colliding hadrons with momenta $p_1$ and $p_2$, $\gamma \gamma$ is the diphoton pair
with invariant mass $\Mgg$, transverse momentum $q_T$ and rapidity $y$, and $X$ is an arbitrary and undetected final state. 

Using the factorization formula, we can write the corresponding fully differential cross section, in $q_T$, $\Mgg$ and $y$, which we denote for simplicity (since our focus is on the $q_T$
distribution) by $d\sigma_{\gg}/dq_T^2$, 
\begin{equation}
\label{dcross}
\f{d\sigma_{\gg}}{d q_T^2}(q_T,\Mgg,s)= \sum_{a,b}
\int_0^1 \!\!\!dx_1 \,\int_0^1 \!\!\!dx_2 \,f_{a/h_1}(x_1,\mu_F^2)
\,f_{b/h_2}(x_2,\mu_F^2) \;
\f{d{\hat \sigma}^{\gg}_{ab}}{d q_T^2}(q_T, \Mgg,y,{\hat s};
\as,\mu_R^2,\mu_F^2)
\end{equation}
(up to power-suppressed corrections), where the $f_{a/h}(x,\mu_F^2)$ ($a=q,{\bar q}, g$)
are the parton densities of the hadron $h$ at the factorization scale $\mu_F$, $\as \equiv \as(\mu_R^2)$,
$d\hat\sigma^{\gg}_{ab}/d{q_T^2}$ is the pQCD \emph{partonic cross section}, 
$s$ ($\hat s = x_1 x_2 s$) 
is the square of the hadronic (partonic) centre--of--mass  energy,  and $\mu_R$ is the renormalization scale.

In the large $q_T$ region ($q_T \sim  \Mgg$) the QCD perturbative series is controlled by a small expansion parameter, 
$\as(\Mgg)$, and a fixed-order calculation of the partonic cross section is theoretically justified. For these values of transverse momentum, 
the QCD radiative corrections are known up to next-to-leading order (NLO)~\cite{Dicus:1987fk,Barger:1989yd,Bern:1994fz,Anastasiou:2002zn}. 

The convergence of the fixed-order perturbative expansion, in the small-$q_T$ region ($q_T\ll \Mgg$), is spoiled by the presence
of powers of large logarithmic terms,  $\as^n\ln^m (\Mgg^2/q_T^2)$.
In order to obtain reliable predictions these terms have to be resummed to all orders.

To apply the transverse momentum resummation, we start by decomposing the partonic cross section as
\begin{equation}
\label{resplusfin}
\f{d{\hat \sigma}^{\gg}_{ab}}{dq_T^2}=
\f{d{\hat \sigma}_{\gg\,ab}^{(\rm res.)}}{dq_T^2}
+\f{d{\hat \sigma}_{\gg\,ab}^{(\rm fin.)}}{dq_T^2}\; .
\end{equation}
The first term on the right-hand side contains all the logarithmically-enhanced contributions,
which have to be resummed to all orders in $\as$,
while the second term
is free of such contributions and can thus be evaluated at fixed order in perturbation theory. 
Using the Fourier transformation between the conjugate variables 
$q_T$ and $b$ ($b$ is the impact parameter),
the resummed component $d{\hat \sigma}^{({\rm res.})}_{\gg\,ab}$
can be expressed as
\begin{equation}
\label{resum}
\f{d{\hat \sigma}_{\gg \,ab}^{(\rm res.)}}{dq_T^2}(q_T,\Mgg,y,{\hat s};
\as,\mu_R^2,\mu_F^2) 
=\f{\Mgg^2}{\hat s} \;
\int_0^\infty db \; \f{b}{2} \;J_0(b q_T) 
\;{\cal W}_{ab}^{\gg}(b,\Mgg,y,{\hat s};\as,\mu_R^2,\mu_F^2) \;,
\end{equation}
where $J_0(x)$ is the $0$th-order Bessel function.
The form factor ${\cal W}^{\gg}$ is best expressed in terms of its \emph{double} Mellin moments ${\cal W}_{N_1 N_2}^{\gg}$, taken with
respect to the variables $z_1, \ z_2$ at fixed $\Mgg$, with
\begin{equation}
z_1 z_2 \equiv z = \frac{\Mgg^2}{\hat{s}}, \quad \frac{z_1}{z_2} = e^{2 y};
\end{equation}
the resummation structure of ${\cal W}_{N_1 N_2}^{\gg}$ can be organized in an exponential 
form~\footnote{For the sake of simplicity we consider here only
the case of  the diagonal terms in the flavour space 
of the partonic indices $a,b$. For a detailed discussion, we refer to Ref.~\cite{Bozzi:2005wk,Bozzi:2007pn}.}
\begin{align}
\label{wtilde}
{\cal W}_{N_1 N_2}^{\gg}(b,\Mgg,y;\as,\mu_R^2,\mu_F^2)
& ={\cal H}_{N_1 N_2}^{\gg}\left(\Mgg, 
\as;\Mgg^2/\mu^2_R,\Mgg^2/\mu^2_F,\Mgg^2/\mures^2
\right) \nonumber \\
&\times \exp\{{\cal G}_{N_1 N_2}(\as,L;\Mgg^2/\mu^2_R,\Mgg^2/\mures^2
)\}
\;\;,
\end{align}
were we have defined the logarithmic expansion parameter $L\equiv \ln (1+{\mures^2 b^2}/{b_0^2})$ \cite{Bozzi:2005wk, Bozzi:2003jy},
and $b_0=2e^{-\gamma_E}$ ($\gamma_E=0.5772...$ 
is the Euler number).

The argument of the logarithmic expansion parameter $L$ ensures the unitary constraint. If we take $b=0$ which directly implies $L=0$, it is easy to demonstrate that after inclusion of the finite component (see Eq.~(\ref{resfin})),  
we exactly recover the fixed-order perturbative value of the total cross section
upon integration of the $q_T$  distribution over $q_T$
(i.e., the resummed terms give a vanishing contribution upon integration over $q_T$). In particular we have,
\begin{align}
\label{restot}
\int_0^{\infty} dq_T^2 \;&
\f{d{\hat \sigma}_{\gg}^{(\rm res.)}}{d q_T^2}(q_T,\Mgg,{\hat s};
\as(\mu_R^2),\mu_R^2,\mu_F^2,\mures^2) 
= \nn \\
\f{\Mgg^2}{\hat s} \;&
{\cal H}^{\gg}\!\left(\Mgg,{\hat s},\as(\mu_R^2);\Mgg^2/\mu^2_R,\Mgg^2/\mu^2_F,\Mgg^2/\mures^2 
\right) \;.
\end{align}

The scale $\mures$ ($\mures\sim \Mgg$), 
which appears on the right-hand side of Eqs.~(\ref{wtilde}) and (\ref{restot}),
is the resummation scale \cite{Bozzi:2005wk}. The form factor ${\cal W}_{N_1 N_2}^{\gg}$ (i.e., the product
${\cal H}_{N_1 N_2}^{\gg} \times \exp\{{\cal G}_{N_1 N_2}\}$) does not depend on $\mures$ when
evaluated to all perturbative orders. Its explicit dependence on $\mures$
appears when ${\cal W}_{N_1 N_2}^{\gg}$ is computed by truncation of the resummed
expression at some level of logarithmic accuracy (see Eq.~(\ref{exponent})
below). 
We can use variations of $\mures$ around $\Mgg$ in order to estimate the size of yet uncalculated 
higher-order logarithmic contributions.

The form factor $\exp\{ {\cal G}_{N_1 N_2}\}$ is universal\footnote{The form factor does not depend on the final state;
all the hard-scattering processes initiated by $q\bar{q}$ ($gg$) annihilation have the same form factor.} and
contains all the terms that order-by-order in $\as$ are logarithmically divergent 
as $b \to \infty$ (or, equivalently, $q_T\to 0$). The resummed logarithmic expansion of the exponent ${\cal G}_{N_1 N_2}$ 
is defined as follows:
\begin{align}
\label{exponent}
{\cal G}_{N_1 N_2}(\as, L;\Mgg^2/\mu^2_R,\Mgg^2/\mures^2)&=L \;g^{(1)}(\as L)+g_{N_1 N_2}^{(2)}(\as L;\Mgg^2/\mu_R^2,\Mgg^2/\mures^2)\nn\\
+ & \f{\as}{\pi} g_{N_1 N_2}^{(3)}(\as L,\Mgg^2/\mu_R^2,\Mgg^2/\mures^2)+\dots
\end{align}
where the term $L\, g^{(1)}$ collects the leading logarithmic (LL) $\mathcal{O}(\alpha_s^{p+n}L^{n+1})$
contributions, the function $g_{N_1 N_2}^{(2)}$ includes
the next-to-leading leading logarithmic (NLL) $\mathcal{O}(\alpha_s^{p+n}L^{n})$ contributions \cite{Kodaira:1981nh}, 
$g_{N_1 N_2}^{(3)}$ controls the NNLL $\mathcal{O}(\alpha_s^{p+n}L^{n-1})$ terms \cite{Davies:1984hs, Davies:1984sp, deFlorian:2000pr,Becher:2010tm}
and so forth; $p$ is the number of powers of $\alpha_s$ in the LO (Born) process.
In Eq.~\eqref{exponent}, $\as L$ is formally of order $1$, so there is
an explicit $\mathcal{O}(\as)$ suppression between different logarithmic orders. The explicit form of the functions
$g^{(1)}$, $g_{N_1 N_2}^{(2)}$ and $g_{N_1 N_2}^{(3)}$ can be found in Ref.~\cite{Bozzi:2005wk}.
The process dependent function ${\cal H}_{N_1 N_2}^{\gg}$ 
does not depend on the impact parameter $b$ and it 
includes all the perturbative
terms that behave as constants as $b \to \infty$. 
It can thus be expanded in powers of $\as$:
\begin{align}
\label{hexpan}
{\cal H}_{N_1 N_2}^{\gg}(\Mgg,\as;\Mgg^2/\mu^2_R,\Mgg^2/\mu^2_F,\Mgg^2/\mures^2)&=
\sigma_{\gg}^{(0)}(\alpha_s, \Mgg)
\Bigl[ 1+\nn \\
\f{\as}{\pi} \,{\cal H}_{N_1 N_2}^{\gg \,(1)}(\Mgg^2/\mu^2_F,\Mgg^2/\mures^2) 
\Bigr. 
+ \Bigl. \left(\f{\as}{\pi}\right)^2 &
\,{\cal H}_{N_1 N_2}^{\gg \,(2)}(\Mgg^2/\mu^2_R,\Mgg^2/\mu^2_F,\Mgg^2/\mures^2)+\dots \Bigr],
\end{align}
where $\sigma_{\gg}^{(0)}$ is the partonic cross section at the Born level. Since
the formalism applies to non strongly-interacting final states, in general the Born cross-section can
only correspond to a $q \bar{q}$ or $gg$ initial state. In the specific case of the diphoton production, both
channels contribute, but at different orders in $\alpha_s$: the $q\bar{q}$ subprocess initiates as a pure QED process
($\mathcal{O}(\alpha_s)^0$), while the $gg$ one requires a fermion loop, starting at $\mathcal{O}(\alpha_s)^2$.

In the present work, we keep contributions up to an uniform order in $\alpha_s$ (and all orders in $\as L$), namely
up to $\as^n L^{n-1}$. For the $q\bar{q}$ channel, this requires the inclusion of the $\mathcal{H}$ coefficients of Eq.~\eqref{hexpan} up to order $2$: 
the first-order coefficients ${\cal H}_{N_1 N_2}^{\gg(1)}$ are known since a long time \cite{deFlorian:2000pr},
while the second-order coefficients ${\cal H}_{N_1 N_2}^{\gg(2)}$ were computed only relatively recently~\cite{Catani:2011qz,Catani:2013tia}.
For the $gg$ channel, it is sufficient to include the leading $\mathcal{H}$ contribution (that is, the Born cross-section) and the
appropriate $\mathcal{G}$ in the exponential of Eq.~\eqref{exponent}. Since it does not require any additional numerical effort,
we decided, in all the plots presented in the proceeding, to include all the terms up to $g_{N_1 N_2}^{(3)}$ in the exponential
$\mathcal{G}$ factor also for this channel. In this way, we technically include some terms which are of higher order in $\alpha_s$ with respect to
those in the $q\bar{q}$ channel; however we checked that those terms 
result in a negligible numerical effect (at 1\% accuracy), that is, the
difference produced by including the higher order terms is within the error bands 
obtained by the scale variations, which verifies the stability of the calculation.

The finite component in the Eq.~(\ref{resplusfin}) is the remaining ingredient of the transverse-momentum cross section.
Since $d\sigma_{\gg}^{({\rm fin.})}$ does not contain large logarithmic terms
in the small-$q_T$ region,
it can be evaluated by truncation of the perturbative series
at a given fixed order.
The finite component is computed starting from the usual
fixed-order perturbative truncation of the partonic cross section and
subtracting the expansion of the resummed part at the same perturbative order.
Introducing the subscript f.o. to denote the perturbative truncation of the
various terms, we have:
\begin{equation}
\label{resfin}
\Bigl[ \f{d{\hat \sigma}_{\gg \,ab}^{(\rm fin.)}}{d q_T^2} \Bigr]_{\rm f.o.} =
\Bigl[\f{d{\hat \sigma}_{\gg \,ab}^{}}{d q_T^2}\Bigr]_{\rm f.o.}
- \Bigl[ \f{d{\hat \sigma}_{\gg \,ab}^{(\rm res.)}}{d q_T^2}\Bigr]_{\rm f.o.} \;.
\end{equation} 
This matching procedure 
between resummed and finite contributions guarantees to achieve uniform theoretical accuracy 
over the region from small to intermediate values of transverse momenta. 
At large values of $q_T$,
the resummation (and matching) procedure is eventually superseded by the
customary fixed-order calculations 
(their theoretical accuracy in the large-$q_T$ region cannot be
improved by resummation of the logarithmic terms that dominate 
in the small-$q_T$ region).

As a summary of our presentation of the transverse momentum resummation formalism we can write that
the inclusion of the functions $g^{(1)}$, $g_{N_1 N_2}^{(2)}$,
${\cal H}_{N_1 N_2}^{\gg (1)}$ in the resummed component,
together with the evaluation of the finite component at LO (i.e. at ${\cal O}(\as)$),
allows us to perform the resummation at NLL+LO accuracy.
This is the theoretical accuracy used in previous studies \cite{Balazs:2007hr,Balazs:2006cc,Nadolsky:2002gj,Balazs:1999yf}
of the diphoton $q_T$ distribution.
If we include also the functions $g_{N_1 N_2}^{(3)}$ and ${\cal H}_{N_1 N_2}^{\gg(2)}$, together 
with the finite component at NLO (i.e. at ${\cal O}(\as^2)$) we can perform calculations at full NNLL+NLO accuracy.

In our particular case, using the ${\cal H}_{N_1 N_2}^{\gg(2)}$ 
coefficient~\cite{Catani:2011qz,Catani:2013tia},
we are thus able to present the complete result for the diphoton $q_T$-distribution up to NNLL+NLO accuracy. The NNLL+NLO (NLL+LO) result includes the {\em full} NNLO (NLO)
perturbative contribution in the small-$\qt$ region.
In particular, the NNLO (NLO) result for the total cross section  
is exactly recovered upon integration
over $q_T$ of the differential cross section $d \sigma_{\gg}/dq_T$ at NNLL+NLO
(NLL+LO) accuracy.

It is known that at small values of $q_T$, the perturbative QCD approach has to be supplemented with non-perturbative contributions, since they become relevant as $q_T$ decreases. A discussion on non-perturbative effects on the $q_T$ distribution is presented in Ref.~\cite{Bozzi:2005wk,Cieri:2015rqa}, and related quantitative results are shown in Sect. \ref{sec:results}.

\section{Numerical results for photon pair production at the LHC}
\label{sec:results}

We are interested in this section in the diphoton production in $pp$ collisions at LHC
energies ($\sqrt{s}=7$~TeV). We present our resummed results at NNLL+NLO accuracy, and compare them with NLL+LO predictions
and with available LHC data~\cite{Aad:2012tba}.
The formulation of the $q_T$ resummation formalism that we use here
is restricted to the production of colourless systems~\footnote{Transverse-momentum resummation for strongly-interacting final states, such as heavy-quark production, has been developed in Refs.~\cite{Li:2013mia,Catani:2014qha}. The first implementation of the $q_T$ subtraction formalism was recently presented for top quark production at hadron colliders~\cite{Bonciani:2015sha}.} $F$, therefore it does not 
treat parton fragmentation subprocesses (here $F$ includes one or two coloured
partons that fragment); for this reason, we concentrate on the direct production of diphotons, and 
we rely on the smooth cone isolation criterion proposed by Frixione~\cite{Frixione:1998jh} (see also 
Ref.~\cite{Frixione:1999gr,Catani:2000jh}) which is defined by requesting
\begin{eqnarray}\label{Eq:Isol_frixcriterion}     
&\sum E_{T}^{had} \leq E_{T \, max}~\chi(r)\;, \;\;\;\;\nonumber\\
&\mbox{inside any} \;\;      
r^{2}=\left( y - y_{\gamma} \right)^{2} +    
\left(  \phi - \phi_{\gamma} \right)^{2}  \leq R^{2}  \;,    
\end{eqnarray}
with a suitable choice for the function $\chi(r)$. This function has to vanish smoothly when its argument goes to zero ($\chi(r) \rightarrow 0 \;,\; \mbox{if} \;\; r \rightarrow 0\,$), and it has to verify \mbox{$\; 0<\chi(r)< 1$}, if \mbox{$0<r<R\,\,.$} One possible choice is
\begin{equation}
\label{Eq:Isol_chinormal}
\chi(r) = \left( \frac{1-\cos (r)}{1-\cos R} \right)^{n}\;,
\end{equation}
where $n$ is typically chosen as $n=1$. This condition implies that, closer to the photon, less hadronic activity is allowed inside the cone. When the parton and the photon are exactly collinear (at $r=0$), the energy deposited inside the cone is required to be exactly equal to zero, and the fragmentation component  (which is a purely collinear phenomenon in perturbative QCD) vanishes completely.

The cancellation of soft gluon effects takes place as in ordinary infrared-safe cross sections, since no region of the phase space is forbidden. That is the main advantage of this criterion: it eliminates all the fragmentation component in an infrared-safe way.
By contrast, it can not be implemented within the usual experimental conditions;
the standard way of implementing isolation in experiments is to use the prescription of Eq.~\eqref{Eq:Isol_chinormal} with a constant
$\chi(r)=1$. In any case, from a purely pragmatic point of view, it has been recently shown~\cite{Butterworth:2014efa} that
if the isolation parameters are tight enough (e.g., $E_{T~max} < 6~$GeV, $R=0.4$), the standard and the smooth cone isolation prescription coincide at the $1\%$ level, which is well within the theoretical uncertainty of our predictions.

The acceptance criteria used in this analysis ($\sqrt{\rm s}=7$~TeV) are those implemented by the ATLAS collaboration analysis \cite{Aad:2012tba};
in  all the numerical results presented in this proceeding, we require $p_T^{\rm harder} \geq 25$~GeV, $p_T^{\rm softer}\geq 22$~GeV, and
we restrict the rapidity of both photons to satisfy $|y_\gamma|<1.37$ and \mbox{$1.52<|y_\gamma| \leq 2.37$}. 
The isolation parameters are set to the values $E_{T~max}=4~$GeV, $n=1$ and $R=0.4$, and the minimum angular separation between the two photons is $R_{\gg}=0.4$. We use the Martin-Stirling-Thorne-Watt (MSTW) 2008 \cite{Martin:2009iq} sets of parton distributions, with
densities and $\as$ evaluated at each corresponding order
(i.e., we use $(n+1)$-loop $\as$ at N$^n$LO, with $n=0,1,2$),
and we consider $N_f=5$ massless quarks/antiquarks and gluons in 
the initial state. The default
renormalization ($\mu_R$) and factorization ($\mu_F$) scales are set to the value
of the invariant mass of the diphoton system,
$\mu_R=\mu_F = M_{\gamma\gamma}$, while the default resummation scale ($\mu_{res}$) is set to $\mu_{res} = \Mgg/2$.
The QED coupling constant $\alpha$ is fixed to $\alpha=1/137$. 
\begin{figure}[htb!]
\begin{center}
\begin{tabular}{cc}
\includegraphics[width=0.474\textwidth]{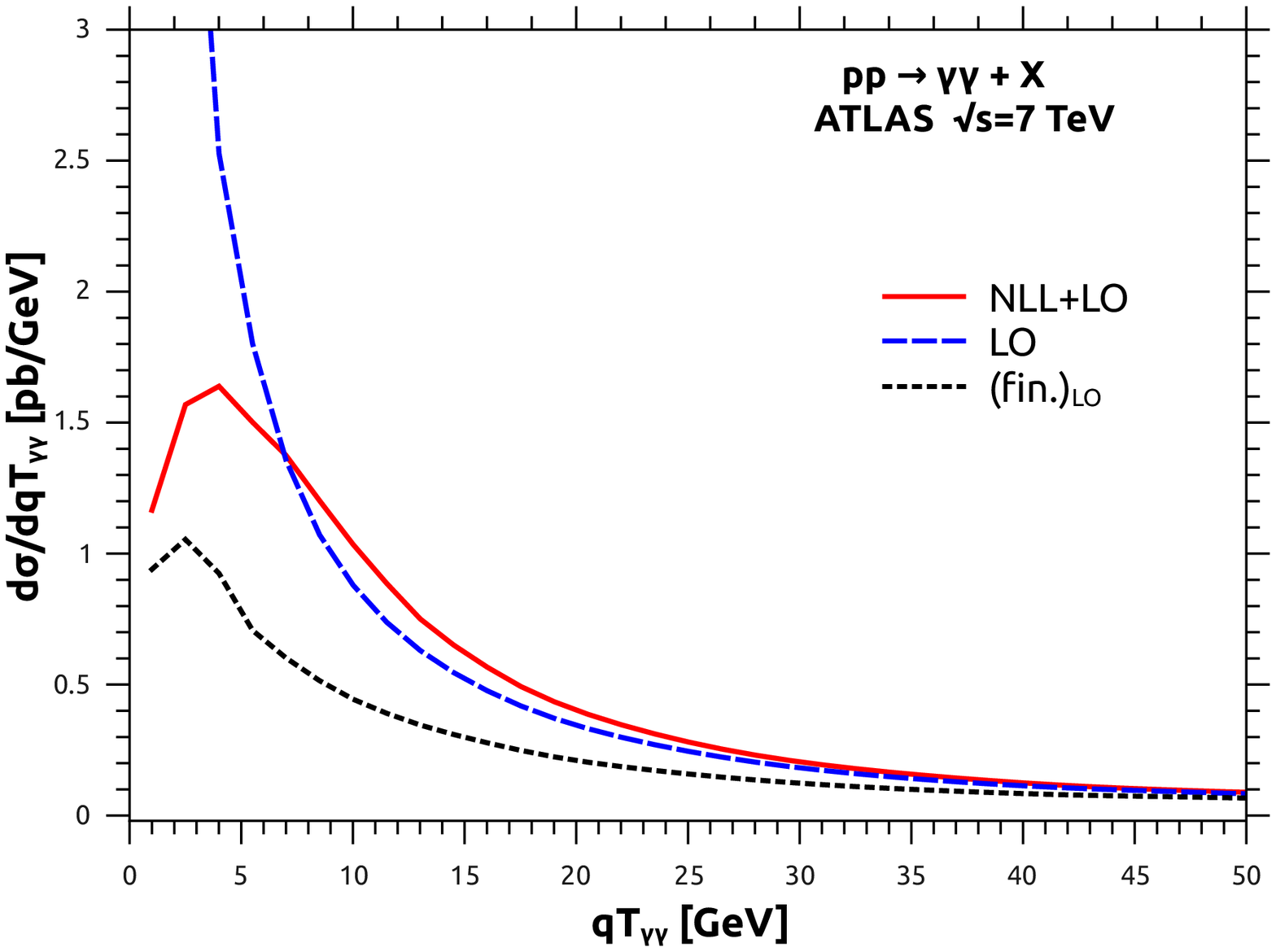} & \includegraphics[width=0.474\textwidth]{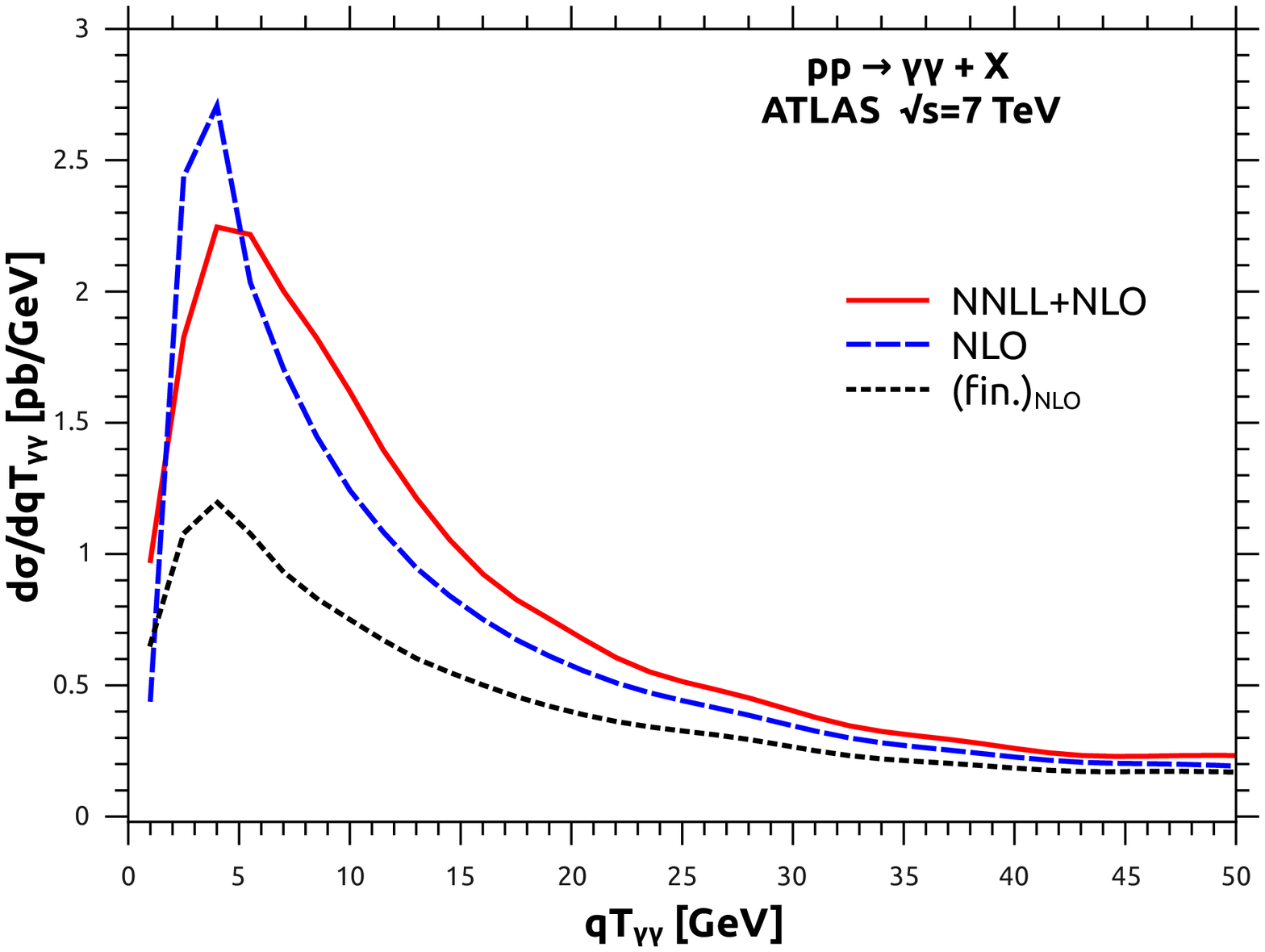}\\
\end{tabular}
\end{center}
\caption{\label{fig1}
 The $q_T$ spectrum of the photon pair (solid red lines) at the LHC (7~TeV): results at NLL+LO 
(left panel) and NNLL+NLO (right panel) accuracy. Each result is compared to the corresponding
fixed-order result (dashed lines) and to the finite component (dotted lines) in Eq.~(\ref{resfin}). 
The resummed
spectrum includes a non-perturbative (NP) contribution parametrized as in Eq.~\eqref{NPcon}.}
\end{figure}

Non-perturbative (NP) effects are expected to be important at very small $q_T$. 
Here we follow the strategy of Ref.~\cite{Bozzi:2005wk}, implementing them
by multiplying  the $b$-space form
factor $\mathcal{W}^{\gg}$ of Eq.~\eqref{resum} by a `NP factor' which consists of a 
gaussian smearing of the form
\begin{equation}
\label{NPcon}
S^a_{NP} = \exp( -C_a \, g_{NP} \, b^2),
\end{equation}
where $a$ denotes the initial state channel, $a=F$ for $q\bar{q}$ and $a=A$ for $gg$ (as usual,
$C_F = (N_c^2 -1)/(2 N_c)$ and $C_A = N_c$).
In order to asses the importance of the NP contributions, in the Ref.~\cite{Cieri:2015rqa} we varied $g_{NP}$ in the interval
from $g_{NP} = 0$~GeV$^2$ (no NP contributions)
to $g_{NP} = 2$~GeV$^2$, corresponding to \textit{moderate} NP effects~\cite{Bozzi:2005wk}. As was shown in Ref.~\cite{Cieri:2015rqa} the NP parameter that shows the better agreement with the data is $g_{NP} = 2$~GeV$^2$, therefore, in the rest of the proceeding we use this value to present all our results.

\begin{figure}[htb]
\begin{center}
\begin{tabular}{cc}
\includegraphics[width=0.474\textwidth]{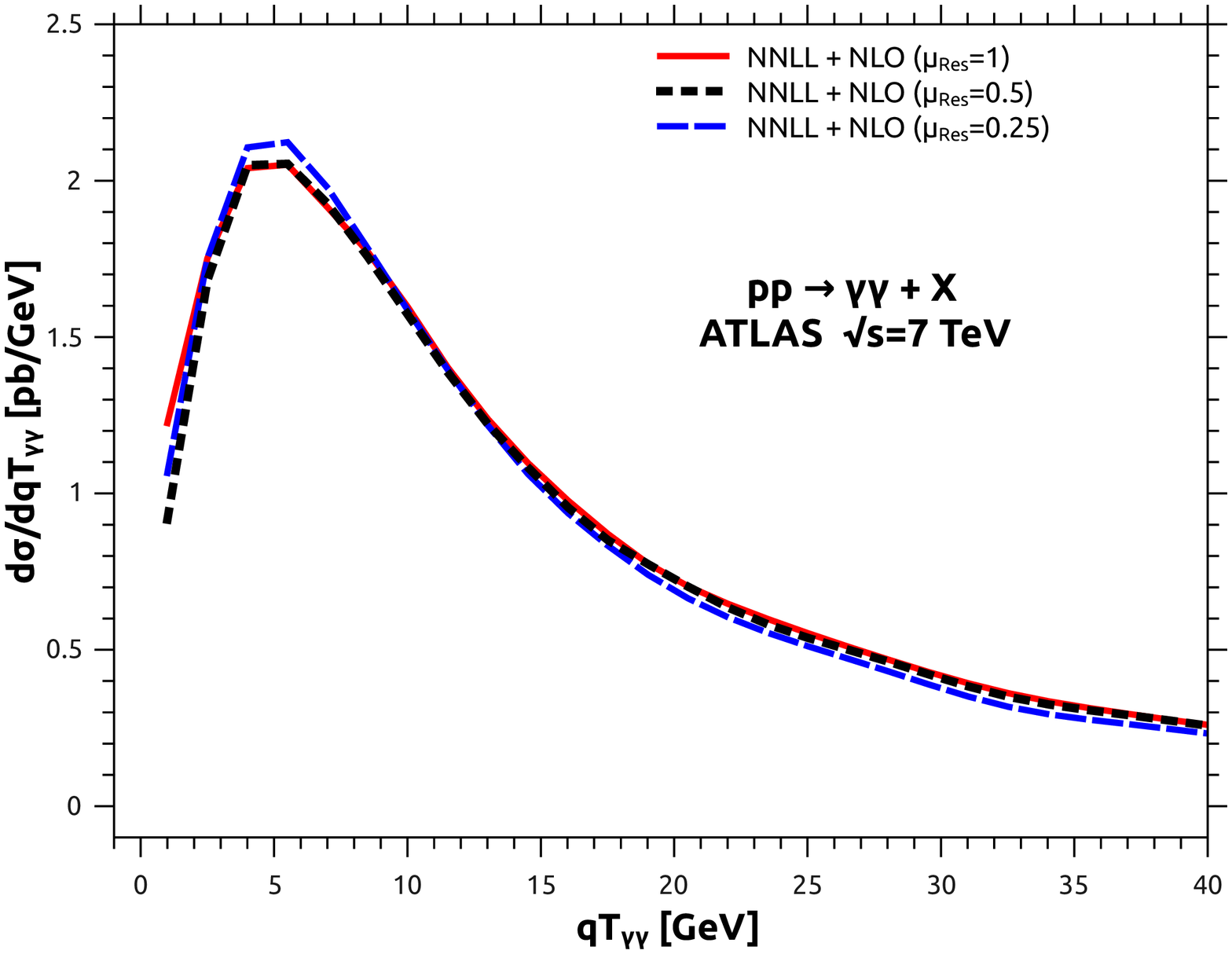} & \includegraphics[width=0.474\textwidth]{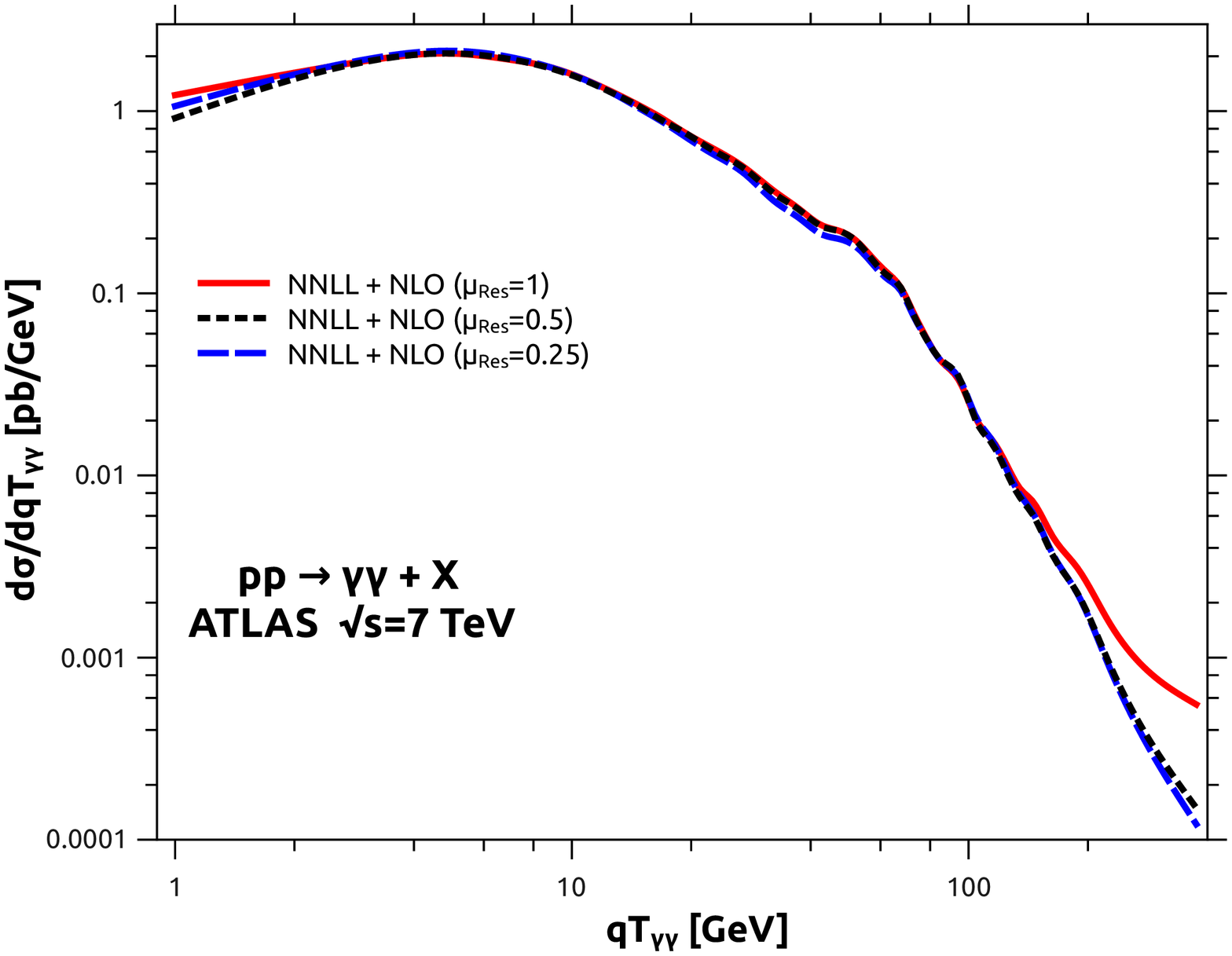}\\
\end{tabular}
\end{center}
\caption{\label{fig2_2}
The $q_T$ spectrum of diphoton production at the LHC ($\sqrt{s}=7$~TeV). Here, the scales 
 $\mu_F=\mu_R=\Mgg$ are kept fixed while we vary the resummation scale $\mu_{res}$ 
 to establish its impact on the cross section. In the left panel we show the 
 range $0$~GeV$<q_{T\gg}<40$~GeV, and in the right panel the full spectra in logarithmic scale.}
\end{figure}

In Fig.~\ref{fig1}, left panel, we present  the NLL+LO $q_T$ spectrum at the LHC ($\sqrt{s}=7$~TeV).
The NLL+LO result (solid line) at the default scales ($\mu_F=\mu_R=\Mgg$;~$\mu_{res}=\Mgg/2$) is 
compared with the
corresponding LO result (dashed line). The LO finite component of the spectrum (see Eq.~(\ref{resplusfin})), is also shown 
for comparison (dotted line).
We observe that the LO result diverges to $+\infty$ as $q_T\to 0$, as expected. The finite component is regular over the full $q_T$ range, it smoothly vanishes as $q_T\to 0$ and gives an important contribution to the NLL+LO result in the low-$q_T$ region.
That is mostly originated by the $qg$ channel, which starts at NLO and provides a {\it subleading} 
correction in terms of logs (single logarithmic terms) but contributes considerably to the cross-section
due to the huge partonic luminosity compared to the formally leading $q\bar{q}$ channel.
The resummation of the small-$q_T$ logarithms
leads to a well-behaved distribution: it vanishes as $q_T\to 0$ and approaches the corresponding LO result
at large values of $q_T$.

The results in the right panel of Fig.~\ref{fig1} are systematically at one order
higher: the $q_T$ spectrum at NNLL+NLO accuracy (solid line) is compared with
the NLO result (dashed line) and with the NLO finite component of the spectrum
(dotted line).
The NLO result diverges to $-\infty$ as $q_T\to 0$ and, at small values of $q_T$,
it has an unphysical peak that is produced by the compensation of negative leading
and positive subleading logarithmic contributions.
The contribution of the NLO finite component to the NNLL+NLO result is of the order of the 50\% at the peak and becomes more important as $q_T$ increases.
A similar quantitative behaviour is observed by considering the contribution of
the NLO finite component to the NLO result. At large values of $q_T$ the contribution of the NLO finite component tends 
to the NLO result. This behaviour indicates that the logarithmic terms are no 
longer dominant and that the resummed
calculation cannot improve upon the predictivity of the fixed-order expansion. We also observe that the position of the peak in
the NNLL+NLO $q_T$ distribution is slightly harder than the corresponding 
NLL+LO $q_T$ distribution. This effect is (in part) due to the large 
transverse-momentum dependence of the fixed order corrections.

The resummed calculation depends on the factorization and 
renormalization scales and on the resummation scale $\mures$, as discussed in Sect.~\ref{sec:theory}. Our convention to compute factorization  and renormalization scale uncertainties is to consider 
independent variations of $\mu_F$ and $\mu_R$ by a factor of two around 
the central values $\mu_F=\mu_R=\Mgg$ in independent way in order to maximise them: ($\mu_F=2~\Mgg$, $\mu_R=\Mgg/2, \mures=\Mgg/2$)
and ($\mu_R=2~\Mgg$, $\mu_F=\Mgg/2, \mures=\Mgg/2$). The uncertainty due to the resummation scale variation
is assessed separately by varying it between $\mu_{res} = \Mgg/4$ and $\mu_{res} = \Mgg$ at fixed $\mu_F$ and $\mu_R$.

\begin{figure}[htb]
\begin{center}
\begin{tabular}{cc}
\includegraphics[width=0.474\textwidth]{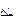} & \includegraphics[width=0.474\textwidth]{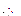}\\
\end{tabular}
\end{center}
\caption{\label{fig3}
 The $q_T$ spectrum of diphoton pairs at the LHC. The
NNLL+NLO result is compared with the NLL+LO result, for the window 0 GeV $< q_T<40$~GeV (left panel)
and the full spectra (right panel). The bands are obtained
by varying $\mu_R$ and $\mu_F$ as explained
in the text.}
\end{figure}


\begin{figure}[htb]
\begin{center}
\begin{tabular}{cc}
\includegraphics[width=0.474\textwidth]{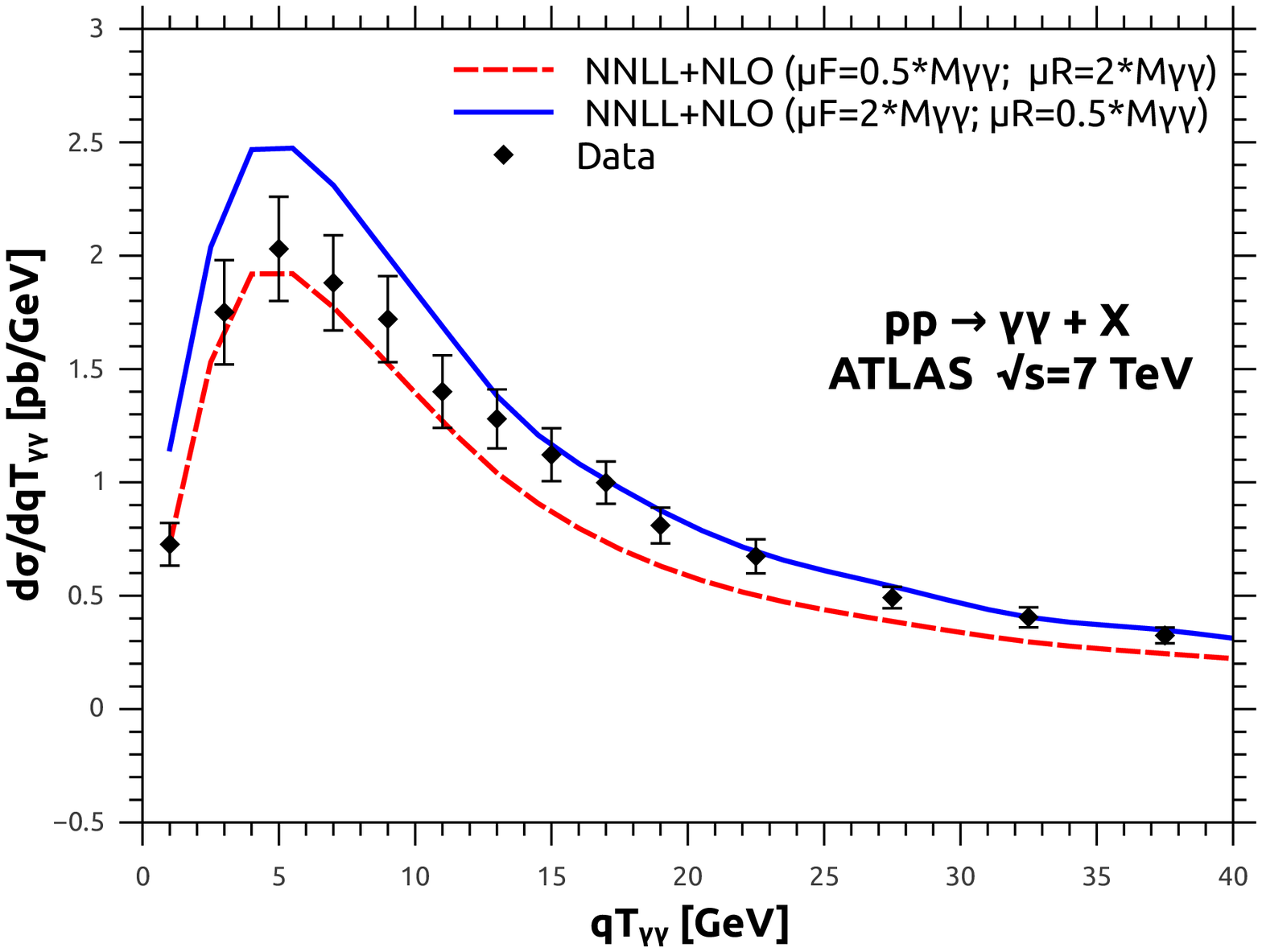} & \includegraphics[width=0.474\textwidth]{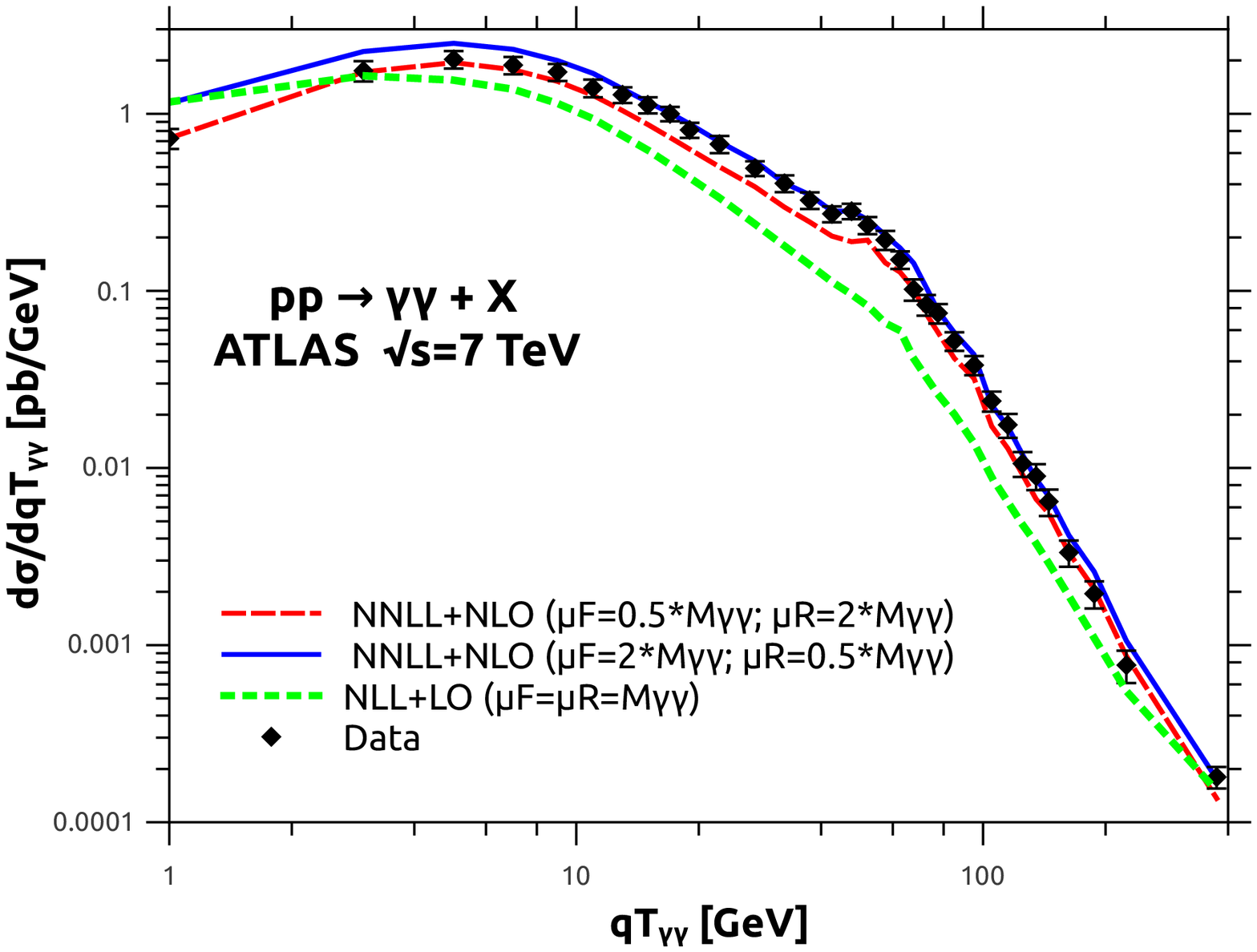}\\
\end{tabular}
\end{center}
\caption{\label{fig5}
 Comparison of the theoretical prediction for the $q_T$ spectrum of diphoton pairs at the LHC with the experimental data. The
NNLL+NLO result is compared with the ATLAS data of Ref.~\cite{Aad:2012tba}, for the window 0 GeV $< q_T<40$~GeV (left panel) and
the full spectra (right panel). In the right panel the NNL+LO distribution at central scale (dotted line) is also shown in order to compare it with the data and the NNLL+NLO result. The bands (solid and dashed lines) are obtained by varying $\mu_R$ and $\mu_F$ as explained
in the text.}
\end{figure}

In order to estimate the impact of the resummation scale $\mures$, we show in Fig.~\ref{fig2_2} the NNLL+NLO transverse momentum distribution for three different 
implementations of the $\mures$ parameter ($\mures=\Mgg/4;\Mgg/2;\Mgg$) at fixed $\mu_F=\mu_R=\Mgg$. The impact of the variation of the $\mures$ scale in the cross section is at per-cent level. In the left panel of Fig.~\ref{fig2_2} we present the transverse momentum distribution for values of $q_T$ within the interval $0$~GeV$<q_T < 40$~GeV, and in the right panel of Fig.~\ref{fig2_2} the full spectra. 
We also notice that the strongest effect of the variation of the $\mures$ scale appears in the
last bin of right panel of Fig.~\ref{fig2_2}. This is expected since the resummation scale effectively {\it sets} the value of transverse momentum at which the logarithms are dominant. A choice of a very large resummation scale affects the distribution at larger transverse momentum and might in general result in a mismatch with the fixed order prediction due to the artificial introduction of unphysically large logarithmic contributions in that region. 

In Fig.~\ref{fig3} we compare the variation of the scales of the NNLL+NLO and NLL+LO predictions, for the interval $0$~GeV$<q_T < 40$~GeV (left panel) and the full spectra (right panel). As in the case of the fixed order calculation~\cite{Catani:2011qz}, the dependence on the scales is not reduced when going 
from NLL+LO to NNLL+NLO. This is mostly because at NNLL+NLO a new channel (gg) opens,  in which 
the box contribution (effectively ``LO'' but formally $\mathcal{O}(\alpha_S^2)$) ruins the reduction of the scale dependence usually expected when adding second order corrections for the $q\bar{q}$ channel and first order corrections for the $qg$ channel. 
Since NNLL+NLO is the first order at which all partonic channels contribute, it is possible to argue that this is the first order at which estimates of theoretical uncertainties through scale variations can be considered as reliable. 

In the right panel of Fig.~\ref{fig3} we observe the so called
\textit{Guillet shoulder}~\cite{Binoth:2000zt}, which is a real radiation 
effect and has its origin in the fixed order contribution. It appears stronger in the 
NNLL+NLO $q_T$ distribution than in the NLL+LO, due to the larger size of the real contributions
at NLO.

\begin{figure}[htb]
\begin{center}
\begin{tabular}{cc}
\includegraphics[width=0.474\textwidth]{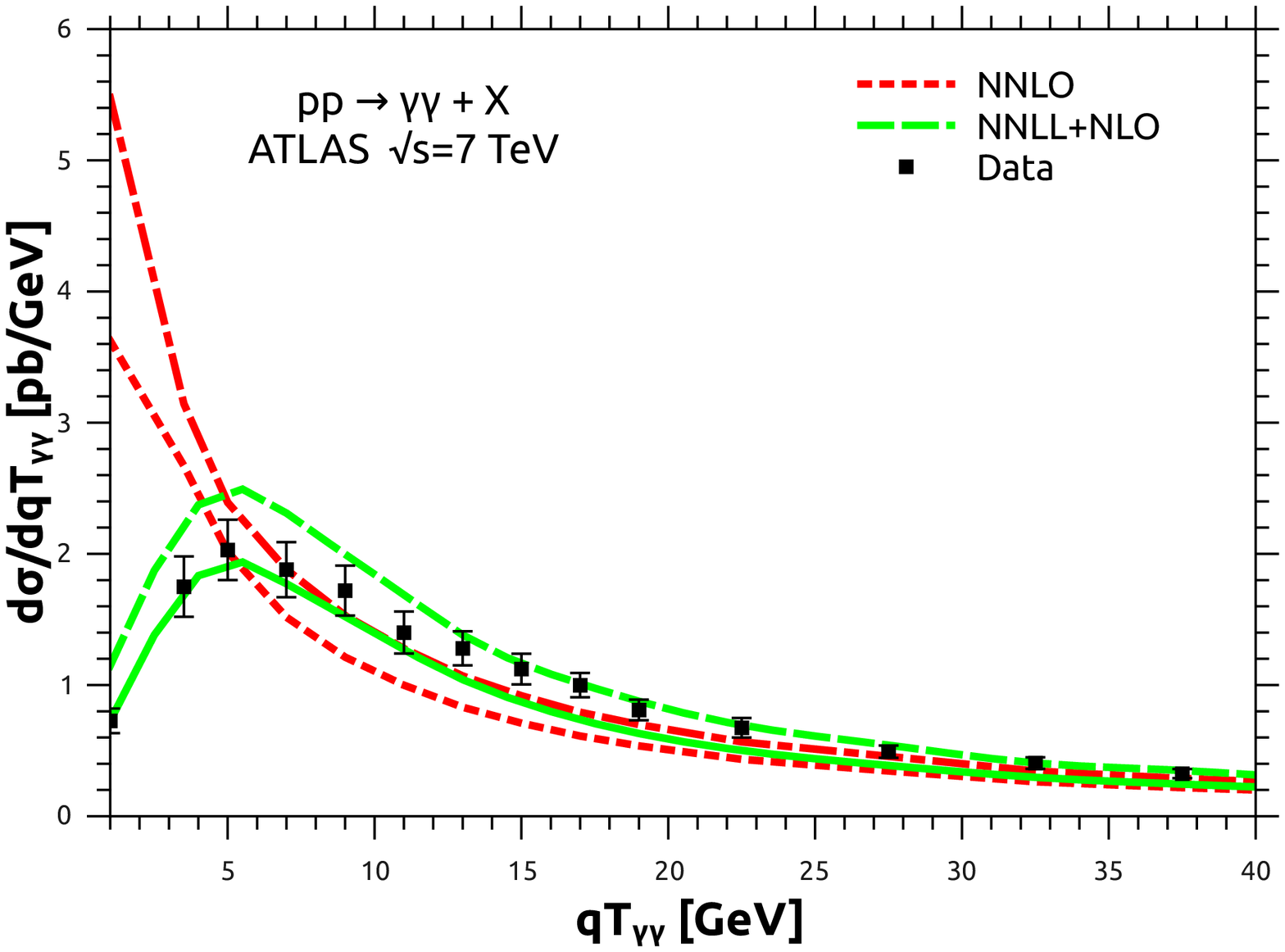} & \includegraphics[width=0.474\textwidth]{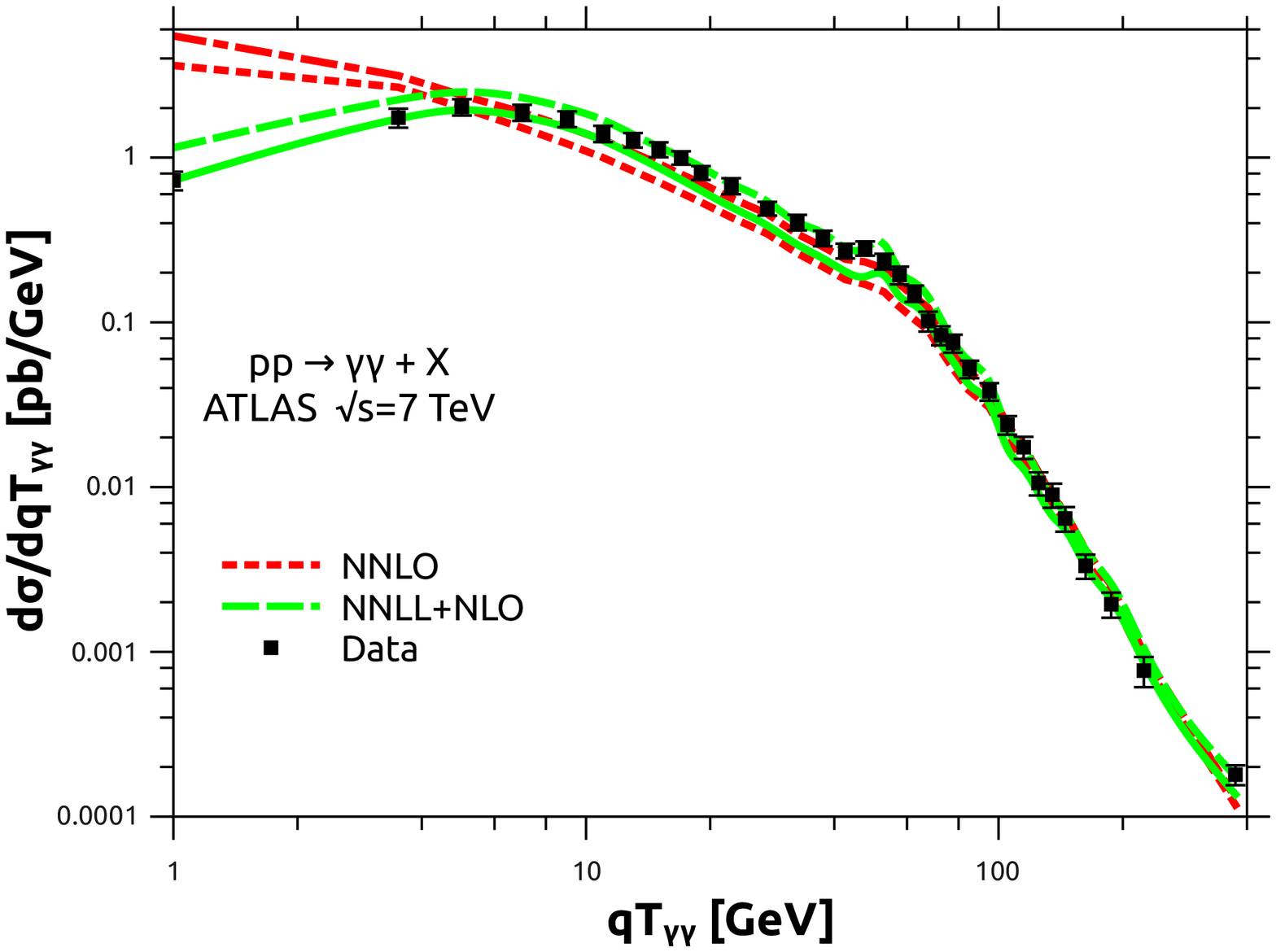}\\
\end{tabular}
\end{center}
\caption{\label{fig6n}
Comparison of the theoretical prediction for the $q_T$ spectrum of diphoton pairs at the LHC with the experimental data. The
NNLL+NLO result (green dashed band) and the fixed order result at NLO (red dotted band) are compared with the ATLAS data of Ref.~\cite{Aad:2012tba}, for the window 0 GeV $< q_T<40$~GeV (left panel) and
the full spectra (right panel). The bands (dotted and dashed lines) are obtained by varying $\mu_R$ and $\mu_F$ as explained
in the text.}
\end{figure}

In Fig.~\ref{fig5} we compare the LHC data ($\sqrt{s}=7$~TeV) from ATLAS~\cite{Aad:2012tba} with our resummed theoretical predictions (at NNLL+NLO and NLL+LO).
We estimate the theoretical uncertainty by the variation of the $\mu_R$ and $\mu_F$ scales.
In the left panel we show the $q_T$ distribution in the window (0~GeV$<q_T < 40$~GeV), 
while in the right ones we show the full spectra in logarithmic scale. 

We observe in general an excellent agreement between the resummed NNLL+NLO prediction and 
the experimental data, that is accurately described within the theoretical uncertainty bands 
in the whole kinematic range. Also we note that the NLL+LO result is not enough to describe 
the phenomenology of the transverse-momentum distribution of the LHC data (Fig.~\ref{fig5}, right panel). By direct comparison to the fixed order prediction, we notice
that the effect of resummation is not only to recover the predictivity of the calculation at
small transverse momentum, but also to improve substantially the agreement with 
LHC data~\cite{Aad:2012tba}.

\begin{figure}[htb]
\begin{center}
\begin{tabular}{cc}
\includegraphics[width=0.474\textwidth]{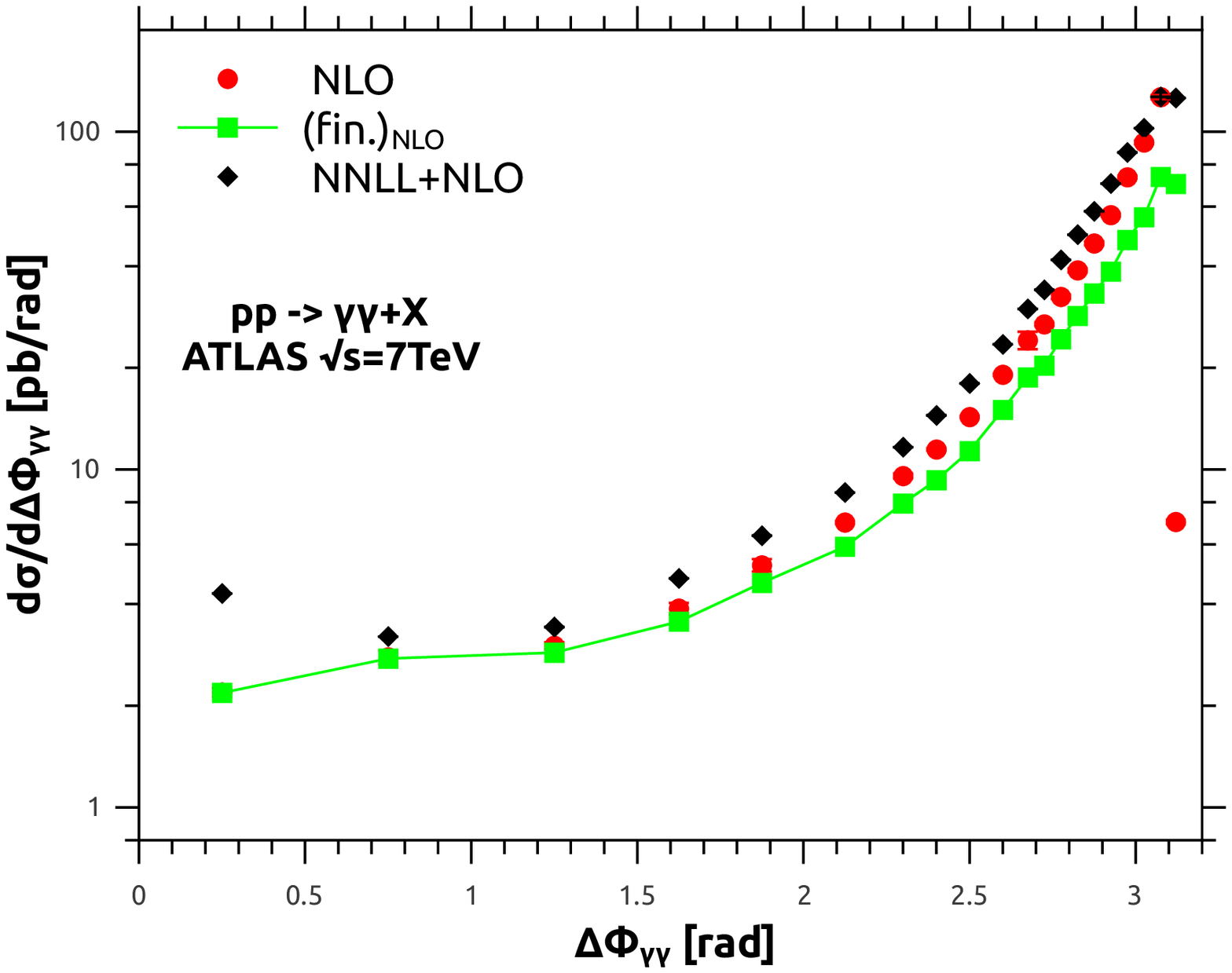} & \includegraphics[width=0.474\textwidth]{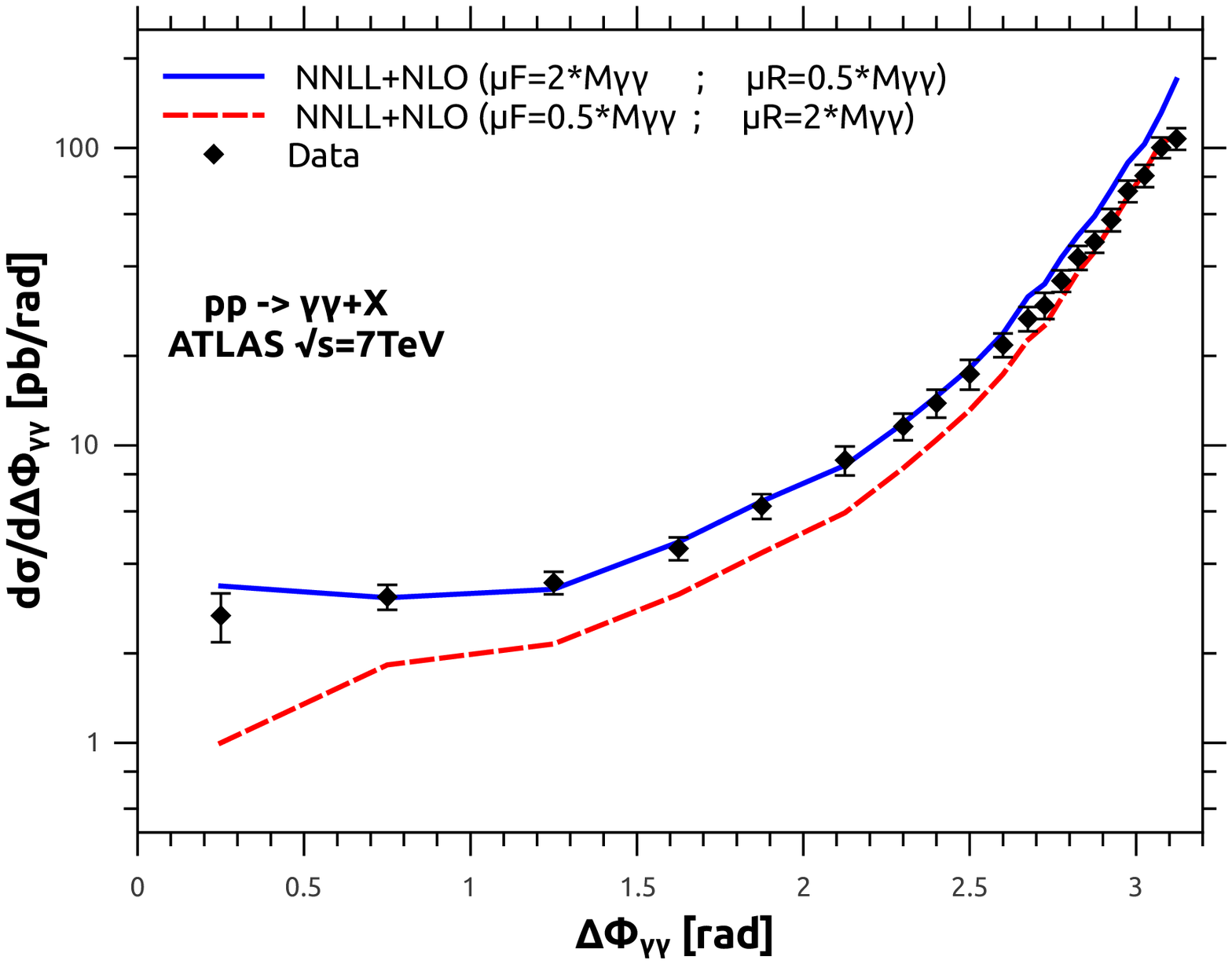}\\
\end{tabular}
\end{center}
\caption{\label{fig6}
 The $\Dpgg$ distribution of diphoton pairs at the LHC. In the left panel, we show the fixed order prediction, the complete resummed prediction
and just the resummed contribution at the central scale. In the right, the full NNLL+NLO result (using
the extreme values for $\mu_R$ and $\mu_F$ to estimate the theoretical uncertainty) is compared with the ATLAS data of Ref.~\cite{Aad:2012tba}.}
\end{figure}

In order to understand how the resummed result improves the fixed order theoretical description of the data, we present in Fig.~\ref{fig6n} the comparison of the theoretical prediction for the $q_T$ spectrum of diphoton pairs at the LHC with the experimental data. The
NNLL+NLO result (green dashed band) and the fixed order result at NLO (red dotted band) are compared with the ATLAS data of Ref.~\cite{Aad:2012tba}, for the window 0 GeV $< q_T<40$~GeV (left panel) and
the full spectra (right panel). Two comments are in order: i) in the small-$q_T$ region ($q_T\ll \mgg$) where the prediction of the fixed order result is not reliable, we observe how the resummed distribution improves the description of the phenomenology of the data; ii) in the large-$q_T$ region ($q_T\sim \Mgg$) where the effects of resummation vanish, the resummed distribution tends to fixed order result, as expected.

While the resummation performed in this work reaches NNLL accuracy formally only for the diphoton transverse momentum distribution, its predictions can be extended to other observables as well, since at least the leading logarithmic contributions have a common origin from soft and collinear emission. 

In Fig.~\ref{fig6} we present the results of the cross section as a function of the azimuthal angle $\Dpgg$. In the left panel we compare the fixed order (NLO), finite (NLO) and full (NNLL+NLO) $\Dpgg$ distributions. The fixed order component dominates
the cross section over the whole $\Dpgg$ range.
However, as could be expected, the effect of resummation is stronger for kinematic configurations
near the $\Dpgg\sim\pi$ which correspond to $q_T\sim0$~GeV. As in the case of the fixed order $q_T$ distribution, the $\Dpgg$ fixed
order differential cross section is not well-behaved near the back-to-back configuration: it actually diverges as
$\Dpgg \rightarrow \pi$ ($q_T \rightarrow 0$). The finite contribution (Eq.~\eqref{resfin}) is well-behaved near the back-to-back configuration, and the full result (NNLL+NLO) improves the description in the region near $\Dpgg\sim 0$. In the right panel of Fig.~\ref{fig6} we compare our theoretical prediction at NNLL+NLO level of 
accuracy (using the variation
of the $\mu_R$ and $\mu_F$ scales to estimate the theoretical uncertainty) with the LHC data~\cite{Aad:2012tba}. We observe that 
the transverse momentum resummation provides a better description of the data with respect to the 
fixed order result. 

\section{Summary}
\label{sec:summa}

We presented the transverse momentum resummation for photon pair production at NNLL
accuracy in hadron collisions. In the small $q_T$ region, we included the resummation of all logarithmically-enhanced perturbative QCD contributions, up to next-to-next-to-leading logarithmic accuracy; at intermediate and large values of $q_T$, we combined the resummation with the fixed next-to-leading order perturbative result. The matching between the fixed-order and the resummed results is performed in such a way as to exactly reproduce the known next-to-next-to-leading order result for the total cross section;
in the end, the calculation consistently includes all perturbative terms up to formal order $\as^2$. The theoretical uncertainty was estimated by varying the various scales (renormalization, factorization and resummation)
introduced by the formalism.
Our results were compared to experimental data, showing good agreement
between theory and experiment over the whole $q_T$ range. With respect
to the fixed-order calculation, the present implementation provides a better description of the data and recovers the correct
physical behaviour in the small $q_T$ region, with the spectrum smoothly going to zero.
The same set-up also allows the calculation of more exclusive observable distributions; the $\Dpgg$ distribution is a given example.

\end{document}